\documentclass[11pt]{article}
\usepackage[margin=1in]{geometry}

\usepackage{amsmath,amsthm,amssymb,hyperref,bm,amsbsy}

\usepackage{bbm}

\usepackage{comment}
\usepackage{xcolor}

\usepackage{threeparttable}
\usepackage{booktabs}
\usepackage{multirow} % display multi row

\usepackage{graphicx}

\usepackage{float}
\usepackage{caption}
\usepackage{subcaption}
\usepackage{epstopdf}
\numberwithin{equation}{section}

\usepackage[ruled,linesnumbered]{algorithm2e}
\RestyleAlgo{ruled}
\SetKwComment{Comment}{/* }{ */}
\let\oldnl\nl% Store \nl in \oldnl
\newcommand{\nonl}{\renewcommand{\nl}{\let\nl\oldnl}}% Remove line number for one line
\SetKwFor{For}{for}{do}{end~for end}

\usepackage{algorithmic}

\usepackage{color,soul}

\usepackage{array}

\hypersetup{
    colorlinks=true, % true for color/false for box
    linkcolor=blue,
    citecolor=red,
    filecolor=magenta,
    urlcolor=blue,
}

\newtheorem{assu}{{\color{black}Assumption}}

\newtheorem{prop}{{\color{blue}Proposition}}

\let\oldproofname=\proofname
\renewcommand{\proofname}{\rm\bf{\oldproofname}}

\DeclareMathOperator*{\argmin}{argmin}

\newcommand{\op}{\operatorname}

\newcommand{\R}{\mathbb{R}}

\newcommand{\bE}{\mathbb{E}}
\newcommand{\dd}{\mathrm{d}}

\title{A deep primal-dual BSDE method for optimal stopping problems}
\author{
Jiefei Yang\thanks{Department of Mathematics, University of Hong Kong, Pokfulam, Hong Kong
  (jiefeiy@connect.hku.hk).}
\and Guanglian Li\thanks{Department of Mathematics, University of Hong Kong, Pokfulam, Hong Kong
  (lotusli@maths.hku.hk).}
}

\begin{document}
\maketitle
\begin{abstract}
We present a new deep primal-dual backward stochastic differential equation framework based on stopping time iteration to 
solve optimal stopping problems. A novel loss function is proposed to learn the conditional expectation, which consists 
of subnetwork parameterization of a continuation value and spatial gradients from present up to the stopping time.
Notable features of the method include: (i) The martingale part in the loss function reduces the variance of stochastic gradients, which facilitates the training of the neural networks as well as alleviates the error propagation of value function approximation; (ii) this martingale approximates the martingale in the Doob-Meyer decomposition, and thus leads to a true upper bound for the optimal value in a non-nested Monte Carlo way. We test the proposed method in American option pricing problems, where the spatial gradient network yields the hedging ratio directly.
\end{abstract}

\noindent\textbf{Keywords:}
Optimal stopping problem; Deep learning; Backward stochastic differential equation; Doob-Meyer decomposition; American option

\section{Introduction}
We study optimal stopping problems of the form $\sup_{\tau \le T} \bE[g(\tau, X_\tau)]$, where the dynamic $(X_t)_{t\in [0,T]}$ is a $d$-dimensional It\^{o} diffusion, $\tau$ is a stopping time, and $g(t, X_t)$ represents the reward at time $t$. Such problems have broad applications in mathematical finance, e.g., pricing and hedging American or Bermudan options. The option holders need to make exercise decision to maximize their rewards. The optimal stopping problem also draws much attention in academia due to its various equivalent forms, including free boundary PDE \cite{wu1997front}, linear complementarity problem \cite{forsyth2002quadratic}, Hamilton-Jacobi-Bellman equation \cite{forsyth2007numerical}, reflected backward stochastic differential equation \cite{el1997reflected}, and backward stochastic differential equation (BSDE) \cite{fujii2015fbsde}. However, most classical methods suffer from the so-called curse of dimensionality, i.e., the computational complexity grows exponentially as the dimension $d$ increases. To tackling this issue, several deep learning-based approaches based on various formulations have been proposed in the past few years, which have demonstrated great success  \cite{sirignano2018dgm, wang2018deep, becker2019deep, hure2020deep, chen2021deep, herrera2021optimal, becker2020pricing, becker2021solving, dong2024randomized}.

The goal of this paper is an efficient and reliable deep learning method for optimal stopping problems. Although the deep neural networks are used as black boxes, the reliability is achieved via computing the true lower and upper bounds. 
Numerically, the continuous-time stopping problem is approximated by the discrete-time one, where we aim to find the optimal stopping time among finitely many opportunities $0 = t_0 < t_1 < \dots < t_N=T$. The optimal value $V_0$ of discrete-time optimal stopping problem is given explicitly by the dynamic programming principle as follows:
\begin{equation} \label{eq:dp-value-iteration}
    \begin{aligned}
        V_N &= g(t_N, X_{t_N}), \\
        V_k &= \max\left( g(t_k, X_{t_k}), \bE[V_{k+1} | X_{t_k}]\right) \quad k = N-1, \dots, 0.
    \end{aligned}
\end{equation}
The method \eqref{eq:dp-value-iteration} is referred to as the value iteration.
Based on the value iteration, accumulation of errors in the approximate value functions has been observed; see, e.g., \cite[Section 5.4]{stentoft2014value} and \cite[Section 5.2]{chen2021deep}. Thus, to achieve convergence, one needs either a sufficient small approximation error for each $V_k$ \cite{hure2020deep} or the monotonicity condition of one-step approximation \cite{guo2015monotone}, which is nontrivial to satisfy in high dimensions. This motivates the use of the stopping time iteration.

Following the celebrated Longstaff and Schwartz method \cite{longstaff2001valuing} and its analysis in \cite{clement2002analysis}, the optimal stopping time $\tau_k := \min\{t_j \ge t_k: V_j = g(t_j, X_{t_j})\}$ is updated iteratively by
\begin{equation} \label{eq:dp-stopping-time}
    \begin{aligned}
        \tau_N &= t_N, \\
        \tau_k &= \begin{cases}
            t_k, \quad &\text{ if } g(t_k, X_{t_k}) \ge \bE[g(\tau_{k+1}, X_{\tau_{k+1}}) | X_{t_k}], \\
            \tau_{k+1}, \quad &\text{ if } g(t_k, X_{t_k}) < \bE[g(\tau_{k+1}, X_{\tau_{k+1}}) | X_{t_k}],
        \end{cases} \quad k = N-1, \dots, 0.
    \end{aligned}
\end{equation}
Although the tower property implies $\bE[V_{k+1} | X_{t_k}] = \bE[g(\tau_{k+1}, X_{\tau_{k+1}}) | X_{t_k}]$, the numerical approximation of the latter has the advantage that the reward function $g$ is known exactly, and thus can avoid the error propagation of approximating $V_k$. The well-known representatives based on stopping time iteration \eqref{eq:dp-stopping-time} and value iteration \eqref{eq:dp-value-iteration} include the Longstaff-Schwartz method \cite{longstaff2001valuing} and the Tsitsiklis and Van Roy method \cite{tsitsiklis2001regression}. A comparison study of these two methods is given in \cite{stentoft2014value}, which shows the superiority of the stopping time iteration.

Numerical methods construct (sub)optimal stopping times resulting in a low biased estimate of the optimal value for the primal problem. Meanwhile, the dual formulation of optimal stopping problems, first introduced in \cite{rogers2002monte, haugh2004pricing}, indicates that the optimal value can be approximated from above using martingales as close as possible to the Doob martingale in Doob-Meyer decomposition. To guarantee the martingale property, one can adopt the nested Monte Carlo simulation \cite{andersen2004primal, becker2019deep}. We summarize the main novelty and contributions as follows. 

\textbf{Contributions}: 
\begin{enumerate}
    \item The key novelty of this work lies in the design of the least squares loss $\mathcal{L}^{\rm bsde}$ (cf. \eqref{eq:novel-loss}) to learn the conditional expectation $\bE[g(\tau_{k+1}, X_{\tau_{k+1}}) | X_{t_k}]$ using neural networks inspired by linear BSDE. We parameterize both continuation value function and its spatial gradient by neural networks. Compared with the projection methods \cite{kohler2010pricing, lapeyre2021neural, becker2020pricing} and the reflected BSDE methods \cite{hure2020deep, chen2021deep}, the loss $\mathcal{L}^{\rm bsde}$ includes an additional martingale part, and hence termed as a deep primal-dual BSDE method.
    \item The martingale part of the loss $\mathcal{L}^{\rm bsde}$ reduces the variance of stochastic gradient estimate compared with the neural network variants of Longstaff and Schwartz method. Variance reduction is crucial for the convergence of stochastic gradient descent-type algorithm \cite{bottou2018optimization}, which is frequently used for training neural networks.
    \item The method allows efficiently computing an upper bound. We use the trained subnetworks of spatial gradients to approximate the Doob martingale directly. This leads to a true upper bound without nested Monte Carlo simulation, avoiding high computational cost especially when the number of time steps $N$ is large. The non-nested method for computing upper bound was proposed in \cite{belomestny2009true}, where a regression estimator is used for the Doob martingale. In the proposed framework, no additional regression is involved since the spatial gradient subnetwork directly induces the martingale.
\end{enumerate}

\textbf{Related works}: Our work is related to the pioneering deep BSDE method \cite{E2017deepbsde} for high-dimensional PDEs, where the BSDE is solved in a forward manner to match the terminal condition at $T$. However, it cannot be directly applied to optimal stopping problems or solving reflected BSDE, as also mentioned in \cite{wang2018deep, hure2020deep, gao2023convergence}. To overcome this issue, several deep learning methods have been proposed recently based on reflected BSDE or value iteration \eqref{eq:dp-value-iteration}  \cite{fujii2019asymptotic, wang2018deep, hure2020deep, chen2021deep, na2023efficient}, and penalized BSDE \cite{peng2024deep}. (Solving the reflected BSDE backwardly corresponds to the value iteration in \eqref{eq:dp-value-iteration}; see  Section \ref{sec:stop-value} for more details.) The convergence of some methods has been analyzed in \cite{hure2020deep} with an {\it a priori} error estimate and in \cite{gao2023convergence} with an {\it a posteriori} error estimate. The proposed deep primal-dual BSDE method differs from the existing ones in that we use the stopping time iteration and thus do not rely on the reflected BSDE. Furthermore, our method is reliable in the sense of small gaps between the true lower and upper bounds.

Several methods for computing the dual upper bound have been developed in the literature, including the primal-dual approach using a simulation within a simulation \cite{andersen2004primal}, pure dual approaches \cite{rogers2010dual, schoenmakers2013dual}, and  the approach using Wiener chaos expansion \cite{lelong2018dual, bayer2023pricing}. The proposed method borrowing ideas of the non-nested method \cite{belomestny2009true}, and involves much lower computational complexity due to no requirement of additional regression.

The deep primal-dual BSDE method is also related to the temporal difference method in reinforcement learning and stochastic control. The least squares loss function for approximating $\bE[g(\tau_{k+1}, X_{\tau_{k+1}})|X_{t_k}]$ can be viewed as the mean squared temporal difference without intermediate rewards, see \cite{zhou2021actor} for a variance reduced least-squares temporal difference method. However, the optimal stopping problem differs from the classical reinforcement learning problem. In particular, the proposed method efficiently approximates the optimal stopping time and also derives an upper bound.

The remainder of this paper is organized as follows. We present in Section \ref{sec:derivation-deep-bsde} the proposed algorithm with the novel loss function and neural network architecture. Using the trained networks, we derive lower and upper bounds of the optimal value $V_0$ in Section \ref{sec:lower-upper-bound}. To demonstrate the advantages of the stopping time iteration \eqref{eq:dp-stopping-time} over the value iteration \eqref{eq:dp-value-iteration} in the approximation of the lower bound, we show in Section \ref{sec:stop-value} that the former is biased low while the latter is biased high in the backward induction. In Section \ref{sec:num}, we evaluate the proposed method on computing lower and upper prices of American/Bermudan options including geometric basket call options, a strangle spread basket option, a put in Heston model, and max-call options. Finally, we conclude in Section \ref{sec:conclusion} with further discussions.

\section{Derivation of the proposed algorithm} \label{sec:derivation-deep-bsde}
%\subsection{Preliminaries on BSDE}\label{ssec:BSDE}
Consider a filtered probability space $(\Omega, \mathcal{F},\mathbb{F}, \mathbb{P})$ with filtration $\mathbb{F} = (\mathcal{F}_t)_{0\le t\le T}$.
Let $(X_t)_{0\le t\le T}$ be a $\R^d$-valued diffusion process satisfying
\begin{equation} \label{eq:Ito-diffusion}
    \dd X_t = \mu(X_t) \,\dd t + \sigma(X_t) \,\dd W_t, \quad X_0 = x_0,
\end{equation}
where $W_t \in \R^d$ is a $\mathbb{P}$-Brownian motion, $\mu(X_t)\in \R^d$, and $\sigma(X_t) \in \R^{d\times d}$.
The finite-horizon continuous-time optimal stopping problem is
\begin{equation} \label{eq:os-original-continuous}
    \sup_{\tau \in \mathcal{T}}\bE[g(\tau, X_\tau)],
\end{equation}
where $g: [0,T]\times \R^d \to \R_+$ is a measurable function and $\mathcal{T}$ is the set of all $\mathbb{F}$-stopping time.
\begin{comment}
We define the optimal value and continuation value at time $t$ by
\begin{equation*}
    V_t = \sup_{\tau\in \mathcal{T}, \tau\ge t}\bE[g(\tau, X_\tau) | \mathcal{F}_t]\quad \text{ and }\quad c(t, X_t) = \sup_{\tau\in \mathcal{T}, \tau> t}\bE[g(\tau, X_\tau) | \mathcal{F}_t],
\end{equation*}
respectively. It follows directly that $V_t = \max(g(t, X_t), c(t, X_t))$.
\end{comment}

To approximate the continuous-time problem numerically, we introduce its time discretization. We take an equidistant time grid $t_k = k\Delta t$ for $k=0,\dots, N$, with $\Delta t = T/N$. The continuous-time problem is approximated by a discrete-time one $V_0 = \sup_{\tau \in \mathcal{J}} \bE[g(\tau, X_{\tau})]$, where $\mathcal{J}$ is the set of all $(\mathcal{F}_{t_k})_{0\le k\le N}$-stopping time. The discrete-time problem can be solved exactly by stopping time iteration in \eqref{eq:dp-stopping-time} or value iteration in \eqref{eq:dp-value-iteration}. The stopping time $\tau_k$ solves the sub-problem $\sup_{\tau\in \mathcal{J}, \tau\ge t_k}\bE[g(\tau, X_\tau)|X_{t_k}]$.

Next, we derive a linear BSDE for the conditional expectation $\bE[g(\tau_{k+1}, X_{\tau_{k+1}})| X_t]$, which represents the continuation value of stopping later than $t$ in the discrete-time problem. For $t < \tau_{k+1}$, the continuation value is given by
\begin{equation} \label{eq:defi-cv-tau}
    c(t, X_t) = \bE[g(\tau_{k+1}, X_{\tau_{k+1}})| X_t].
\end{equation}
By Feynman-Kac formula, it satisfies a parabolic PDE
\begin{equation} \label{eq:cv-parabolic-pde}
    \frac{\partial c}{\partial t}(t, x) + \mu(x)^\top \nabla_x c(t, x) + \frac{1}{2}\op{Tr}\left( \sigma(x)\sigma(x)^\top \op{Hess}_x c(t, x) \right) = 0,
\end{equation}
for $t\le \tau_{k+1}$. Applying It\^{o}'s formula to $c(t, X_t)$, we get
\begin{equation} \label{eq:cv-ito-formula}
    \begin{aligned}
        c(t, X_t) &= g(\tau_{k+1}, X_{\tau_{k+1}}) - \int_t^{\tau_{k+1}} \nabla_x c(s, X_s)^\top \sigma(X_s) \, \dd W_s \\
        &\quad - \int_t^{\tau_{k+1}} \left( \frac{\partial c}{\partial s}(s, X_s) + \mu(X_s)^\top \nabla_x c(s, X_s) + \frac{1}{2}\op{Tr}\left( \sigma(X_s)\sigma(X_s)^\top \op{Hess}_x c(s, X_s) \right) \right) \,\dd s.
    \end{aligned}
\end{equation}
This, combined with the PDE \eqref{eq:cv-parabolic-pde}, leads to a linear BSDE
\begin{equation} \label{eq:cv-linear-bsde}
    c(t, X_t) = g(\tau_{k+1}, X_{\tau_{k+1}}) - \int_t^{\tau_{k+1}} \nabla_x c(s, X_s)^\top \sigma(X_s) \, \dd W_s.
\end{equation}
Here, the It\^{o} integral $\int_t^{\tau_{k+1}}\nabla_x c(s, X_s)^\top \sigma(X_s) \, \dd W_s$ is a martingale when $\bE[|\nabla_x c(s, X_s)^\top \sigma(X_s)|^2] < \infty$ for $t \le s \le \tau_{k+1}$. Then \eqref{eq:cv-linear-bsde} is a precise form of martingale representation.

Below we will abbreviate the notation $X_{t_k}$ to $X_k$ for $k=0, \dots, N$. Let $\hat{c}_k(X_k)$ be the approximate continuation value at time $t_k$. The Euler--Maruyama approximation of \eqref{eq:cv-linear-bsde} gives
\begin{equation} \label{eq:cv-discrete-bsde}
    \hat{c}_k(X_k) = g(\tau_{k+1}, X_{\tau_{k+1}}) - \sum_{j=k}^{\tau_{k+1}/\Delta t - 1} \nabla \hat{c}_j(X_j)^\top \sigma(X_j) \Delta W_j,\quad 
   \mbox{with }\Delta W_j := W_{t_{j+1}} - W_{t_j}.
\end{equation}

\subsection{A novel loss function}
The proposed algorithm solves the optimal stopping problem by updating the stopping time iteratively by \eqref{eq:dp-stopping-time}, where we approximate $\hat{c}_k$ by one neural network $\mathcal{C}_{k,\theta}$ and the gradient $\nabla \hat{c}_k$ by another neural network $\mathcal{G}_{k, \theta}$, with all the trainable parameters involved in these two neural networks denoted as $\theta$. Inspired by \eqref{eq:cv-discrete-bsde}, the neural networks are trained by minimizing the mean squared loss function
\begin{equation} \label{eq:noval-loss-init}
    \mathcal{L}^{\rm bsde} = \bE\left[ \left( \mathcal{C}_{k,\theta}(X_k) - g(\tau_{k+1}, X_{\tau_{k+1}}) + \sum_{j=k}^{\tau_{k+1}/\Delta t - 1}\mathcal{G}_{j,\theta}(X_j)^\top \sigma(X_j)\Delta W_j \right)^2 \right].
\end{equation}
Note that when step $t_{k+1}$ is finished, the neural networks $(\mathcal{G}_{j,\theta})_{j\ge k+1}$ have already been trained successfully. Thus, at step $t_k$, we only need to minimize the following equivalent loss:
\begin{equation} \label{eq:novel-loss}
    \mathcal{L}^{\rm bsde} = \bE\left[ \left( \mathcal{C}_{k,\theta}(X_k) - g(\tau_{k+1}, X_{\tau_{k+1}}) + \mathcal{G}_{k,\theta}(X_k)^\top \sigma(X_k)\Delta W_k + \sum_{j=k+1}^{\tau_{k+1}/\Delta t - 1}\Delta M_j \right)^2 \right],
\end{equation}
where $\Delta M_j := \mathcal{G}_{j,\theta}(X_j)^\top \sigma(X_j)\Delta W_j$ and we utilize $(\Delta M_j)_{j\ge k+1}$ computed in previous steps.

Using neural network to approximate continuation value is not new in the literature. The neural network variants of Longstaff and Schwartz method \cite{kohler2010pricing, lapeyre2021neural, becker2020pricing} compute the continuation value by minimizing a mean squared loss of the form
\begin{equation} \label{eq:longstaff-schwartz-loss}
    \mathcal{L}^{\rm ls} = \bE\left[ \left( \mathcal{C}_{k,\theta}(X_k) - g(\tau_{k+1}, X_{\tau_{k+1}})\right)^2 \right].
\end{equation}
Compared with the loss $\mathcal{L}^{\rm ls}$, the advantages of the new loss $\mathcal{L}^{\rm bsde}$ in \eqref{eq:novel-loss} 
are two-fold. First, the spatial gradient networks $\mathcal{G}_{k,\theta}$ allow constructing the Doob martingale, which 
naturally induces an upper bound for the optimal value $V_0$; see Section \ref{subsec:upper-bound} below for details.
Second, we can achieve higher accuracy using the loss $\mathcal{L}^{\rm bsde}$ in \eqref{eq:novel-loss} in training due 
to the low variance of its stochastic gradient. Specifically, during the training, a random batch of samples is used to 
approximate the gradient $\nabla_\theta\mathcal{L}^{\rm bsde}$. For the loss $\mathcal{L}^{\rm ls}$, one random drawn 
stochastic gradient is, for $\omega \in \Omega$, given by
\begin{equation} \label{eq:stochastic-grad-estimate-ls}
    \hat{G}^{\rm ls}\left(\omega, X_k, \tau_{k+1}, X_{\tau_{k+1}}\right) = 2\left( \mathcal{C}_{k,\theta}(X_k(\omega)) - g(\tau_{k+1}(\omega), X_{\tau_{k+1}(\omega)}) \right) \partial_\theta \mathcal{C}_{k,\theta}(X_k(\omega)),
\end{equation}
which serves as an estimator of the true gradient of \eqref{eq:longstaff-schwartz-loss} with respect to the network parameter $\theta$. However, even if $\mathcal{C}_{k, \theta}$ is the exact continuation value, the gradient estimator \eqref{eq:stochastic-grad-estimate-ls} can be far away from zero leading to large variance, while the true gradient $\nabla_\theta \mathcal{L}^{\rm ls}$ of the loss at an optimal point $\theta^*$ should be zero. In comparison, using the proposed loss $\mathcal{L}^{\rm bsde}$ in \eqref{eq:noval-loss-init} or \eqref{eq:novel-loss}, one randomly drawn stochastic gradient with respect to $\theta$ in $\mathcal{C}_{k,\theta}$ is, for $\omega \in \Omega$,
\begin{equation} \label{eq:stochastic-grad-estimate-bsde}
\begin{aligned}
     &\quad \hat{G}^{\rm bsde}\left(\omega, X_k, \tau_{k+1}, X_{\tau_{k+1}}, \left(\Delta W_j\right)_{j=k}^{\tau_{k+1}/\Delta t-1}\right) \\
     &= 2\bigg(\mathcal{C}_{k,\theta}(X_k(\omega)) - g(\tau_{k+1}(\omega), X_{\tau_{k+1}(\omega)}) + \sum_{j=k}^{\tau_{k+1}(\omega)/\Delta t - 1} \mathcal{G}_{j,\theta}(X_j(\omega))^\top \sigma(X_j(\omega))\Delta W_j(\omega) \bigg) \partial_\theta \mathcal{C}_{k,\theta}(X_k(\omega)).
\end{aligned}
\end{equation}
Near the optimal point $\theta^*$, one can expect the gradient estimator \eqref{eq:stochastic-grad-estimate-bsde} to be almost vanishing due to the martingale term. Therefore, the proposed loss function $\mathcal{L}^{\rm bsde}$ has lower variance for stochastic gradients, which is summarized in the following proposition.

\begin{prop}[First and second moments of stochastic gradients] \label{prop:var-stochastic-gradient}
Let $\mathcal{C}_{k,\theta}$ be the exact continuation value at $t_k$. Let $C := 4\bE[\partial_\theta \mathcal{C}_{k,\theta}(X_k)^2]$ be a constant and $T>0$ be the finite time horizon. Then for the first moments of \eqref{eq:stochastic-grad-estimate-ls} and \eqref{eq:stochastic-grad-estimate-bsde}, we have
\begin{equation*}
    \bE[\hat{G}^{\rm ls}] = \bE[\hat{G}^{\rm bsde}].
\end{equation*}
Furthermore, the second moment of the stochastic gradient \eqref{eq:stochastic-grad-estimate-ls} satisfies
\begin{equation} \label{eq:variance-sg-ls}
    \bE\left[(\hat{G}^{\rm ls})^2 \right] = C \bE\left[ \int_{t_k}^{\tau_{k+1}} \left( \nabla_x c(s, X_s)^\top \sigma(W_s)\right)^2\,\dd s \right] ,
\end{equation}
while the second moment of the stochastic gradient \eqref{eq:stochastic-grad-estimate-bsde} satisfies
\begin{equation} \label{eq:variance-sg-bsde}
    \bE\left[ (\hat{G}^{\rm bsde})^2 \right] \le CT \sup_{t_j\le s\le t_{j+1}, j\ge k} \bE\left[ \left|\nabla_x c(s, X_s)^\top \sigma(X_s) - \mathcal{G}_{j,\theta}(X_j)^\top \sigma(X_j)\right|^2 \right].
\end{equation}
\end{prop}

Estimate \eqref{eq:variance-sg-bsde} implies that a good network $\mathcal{G}_{j,\theta}$ approximation to $\nabla_x c(t_k, X_k)$ and sufficiently small step size $\Delta t$ lead to small variance of stochastic gradients in the BSDE formulation \eqref{eq:stochastic-grad-estimate-bsde}, which, however, is not true for the squared loss formulation \eqref{eq:longstaff-schwartz-loss}. This small variance of gradient estimate is crucial in training neural networks using stochastic gradient descent-type algorithms, e.g., Adam optimizer. Indeed, the stochastic gradient method suffers from adverse effect of noisy gradient, which impacts the convergence \cite[Section 4.5]{bottou2018optimization}. The small variance of $\hat{G}^{\rm bsde}$ indicates that the proposed loss function $\mathcal{L}^{\rm bsde}$ has less noisy gradient estimates.

\subsection{Neural network architecture and training} \label{subsec:nn-architecture-training}
Minimizing the loss function \eqref{eq:novel-loss} amounts to finding the optimal parameters $\theta^*$ in the value 
subnetwork $\mathcal{C}_{k,\theta}$ and the gradient subnetwork $\mathcal{G}_{k,\theta}$. We use the standard feedforward 
neural network (FNN) with $N_{\rm layer}-1$ hidden layers of the form
\begin{equation*}
    x\in \R^{d_{\rm in}} \mapsto A_{N_{\rm layer}}\circ \sigma \circ A_{N_{\rm layer}-1}\dots \circ \sigma \circ A_1(x) \in \R^{d_{\rm out}},
\end{equation*}
where $A_\ell$, $\ell = 1, \dots, N_{\rm layer}$, are affine transformations, $\sigma = \max(\cdot, 0)$ is the ReLU activation function applied componentwise, and $d_{\rm in}$ and $d_{\rm out}$ are respectively input and output dimensionality. Suppose that the $\ell$-th hidden layer has $d_\ell$ neurons for $\ell = 1,\dots, N_{\rm layer}-1$. Then the dimension of FNN parameter space is
\begin{equation*}
    (d_{\rm in}\times d_1 + d_1) + (d_1 \times d_2 + d_2) + \dots + (d_{N_{\rm layer}-1}\times d_{\rm out} + d_{\rm out}), \quad \text{ for } N_{\rm layer}\ge 2.
\end{equation*}
In view of \eqref{eq:cv-discrete-bsde} and \eqref{eq:novel-loss}, since $\mathcal{C}_{k,\theta}$ approximates the continuation value $\hat{c}_k$ and $\mathcal{G}_{k,\theta}$ approximates its gradient $\nabla \hat{c}_k$, $\mathcal{C}_{k,\theta}$ has one output and $\mathcal{G}_{k,\theta}$ has $d$ outputs.

\begin{figure}[htbp!]
    \centering
    \includegraphics[width=0.5\linewidth]{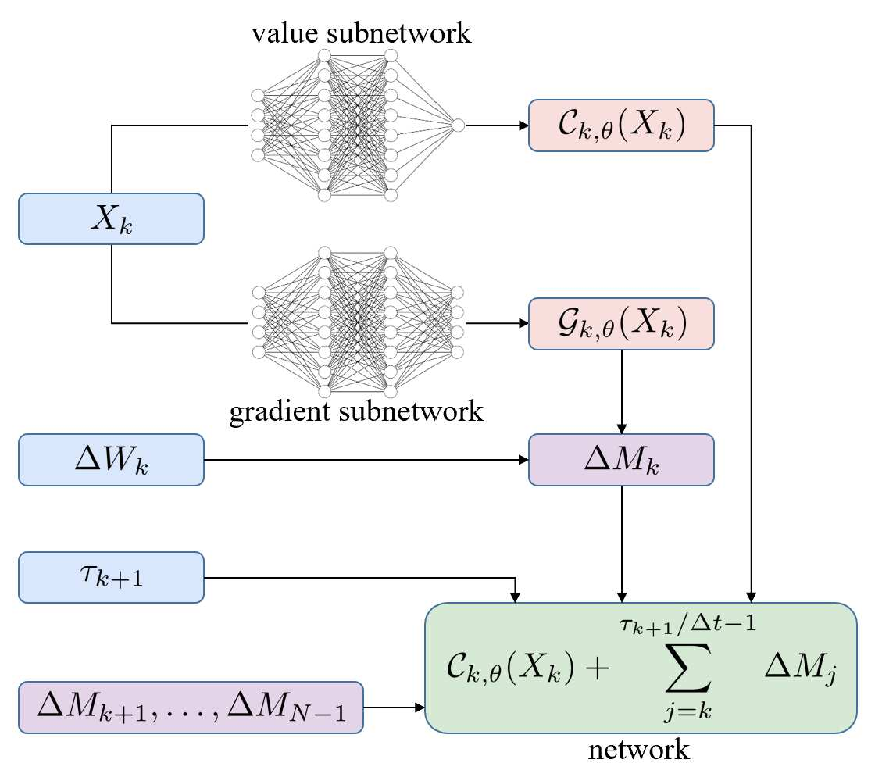}
    \caption{Flowchart of the network for the time step $t_k$.}
    \label{fig:flowchart_network}
\end{figure}

The neural network structure for the time step $t_k$ is illustrated in Figure \ref{fig:flowchart_network}. After assembling the network, the loss function $\mathcal{L}^{\rm bsde}$ in \eqref{eq:novel-loss} is the mean squared error between the neural network prediction and the future optimal reward $g(\tau_{k+1}, X_{\tau_{k+1}})$. We apply mini-batch stochastic gradient descent-type algorithm such as Adam \cite{kingma2015adam} to train the neural network. The pseudo-code of the proposed deep primal-dual BSDE method is summarized in Algorithm \ref{alg:deepBSDEOS}, where $N_{\rm epochs}$ is the number of epochs and $N_{\rm steps}$ is the number of steps per epoch for training.

\begin{algorithm}[htbp!]
\caption{Training deep primal-dual BSDE network for optimal stopping}
\label{alg:deepBSDEOS}
\begin{algorithmic}[1]
\REQUIRE Parameters of the optimal stopping problem and hyperparameters for training.
\ENSURE Trained networks $\mathcal{C}_{k,\theta}$, $\mathcal{G}_{k,\theta}$ for $k=1,\dots, N-1$.
\STATE Generate samples $(X_k)_{k=0}^N$, $(\Delta W_k)_{k=0}^{N-1}$ of forward SDEs \eqref{eq:Ito-diffusion}.
\STATE Initialize $\tau_N = T$.
\FOR {$k = N-1:-1:1$}
    \STATE Initialize the network for time step $t_k$.
    \FOR {$\text{epoch} = 1:N_{\rm epochs}$}
        \FOR {$\text{step} = 1:N_{\rm steps}$}
            \STATE Update network parameters by minimizing the loss \eqref{eq:novel-loss}.
        \ENDFOR
    \ENDFOR
    \STATE Evaluate $\mathcal{C}_{k,\theta}(X_k)$ and $\Delta M_k$ using the trained network.
    \STATE Update $\tau_k$ and $g(\tau_k, X_{\tau_k})$.
\ENDFOR
\end{algorithmic}
\end{algorithm}

We now consider several practical strategies for facilitating the training of the neural networks. First, as observed in \cite{kohler2010pricing, becker2019deep, becker2021solving, chen2021deep}, including the reward function $g(t, x)$ as an input feature improves the efficiency in the training process. In many applications (see Section \ref{sec:num}), the reward function $g(t, x)$ is nonnegative with the form $g(t, x) = \max(\phi(t, x), 0)$. We use the version of \cite{chen2021deep}, where $\phi(t_k, \cdot)$ is considered as an input feature of the value subnetwork $\mathcal{C}_{k,\theta}$. This choice is motivated by the observation that the reward can be recovered by applying the ReLU activation function on the input $\phi(t_k, \cdot)$. For the gradient subnetwork $\mathcal{G}_{k,\theta}$, we observe from experiments that including the feature $\phi(t_k, \cdot)$ will not improve the performance. Thus, we only use $X_k$ as input features for $\mathcal{G}_{k,\theta}$. Second, 
$\nabla_x c(t, x)$ can have steep slopes along some directions when $t \to T^-$ since $g(T, x)$ may not be differentiable on the whole domain in applications, which poses challenges in neural network approximation. Thus, we double the number of epochs for training in the time step $t_{N-1}$ compared with other time steps. Last, a good initialization can improve the efficiency of training. We apply Xavier initialization \cite{glorot2010understanding} in the time step $t_{N-1}$. For $t_k$ with $k\le N-2$, $\mathcal{C}_{k,\theta}$ and $\mathcal{G}_{k,\theta}$ are initialized as the trained subnetworks $\mathcal{C}_{k+1,\theta}$ and $\mathcal{G}_{k+1,\theta}$, respectively,
because the continuation value function $c(t, x)$ evolves gradually from $g(T, x)$ backward in time.

\section{Lower and upper bounds} \label{sec:lower-upper-bound}
In this section, we derive lower and upper bounds for the optimal value $V_0 = \sup_{\tau \in \mathcal{T}} \bE[g(\tau, X_\tau)]$ using the trained networks $\mathcal{C}_{k,\theta}$ and $\mathcal{G}_{k,\theta}$. The lower bound is derived since computing the stopping time iteration \eqref{eq:dp-stopping-time} numerically gives a suboptimal stopping rule. The dual upper bound can be calculated from the gradient network $\mathcal{G}_{k,\theta}$, where the Doob martingale is approximated by an It\^{o} integral directly without using additional nested Monte Carlo sampling.

\subsection{Lower bound}
Using the trained network $\mathcal{C}_{k,\theta}$, the approximate stopping time $\hat\tau_k$ is updated by
\begin{equation*}
    \begin{aligned}
        \hat{\tau}_N &= t_N, \\
        \hat{\tau}_k &= \begin{cases}
            t_k, \quad &\text{ if } g(t_k, X_k) \ge \mathcal{C}_{k,\theta}(X_k) \\
            \hat{\tau}_{k+1}, \quad &\text{ otherwise }
        \end{cases} \quad k = N-1, \dots, 0.
    \end{aligned}
\end{equation*}
This represents a suboptimal stopping time to the original problem. Thus, it gives a lower bound $L = \bE[g(\hat{\tau}_0, X_{\hat{\tau}_0})] \le V_0$. We use Monte Carlo approximation for $L$. First, we sample $N_L$ independent paths $(X_k^{(i)})_{k=0}^N$, $i=1,2,\dots,N_L$, for the stochastic process $(X_k)_{k=0}^N$. It has been pointed out in \cite{glasserman2004monte} that using the same path for decision making and valuation may introduce bias. To ensure that we calculate a reliable lower bound, the sample paths $(X_k^{(i)})_{k=0}^N$ should be independent of the paths used for training the network. Therefore, we derive an unbiased estimate
\begin{equation*}
    \hat{L} = \frac{1}{N_L} \sum_{i=1}^{N_L} g\left(\hat{\tau}_0^{(i)}, X_{\hat{\tau}_0^{(i)}} \right).
\end{equation*}
for the lower bound $L$. The algorithm for estimating the lower bound is given in Algorithm \ref{alg:lower-deepBSDEOS}.

\begin{algorithm}[htbp!]
\caption{Computing lower bound}
\label{alg:lower-deepBSDEOS}
\begin{algorithmic}[1]
\REQUIRE Trained network $\mathcal{C}_{k,\theta}$, $k=1,\dots, N-1$. Sample size $N_L$.
\ENSURE Lower bound estimate $\hat{L}$, confidence interval
\STATE Generate $N_L$ samples $(X_k^{(i)})_{k=0}^N$, $i=1,\dots, N_L$, of forward SDEs \eqref{eq:Ito-diffusion}.
\STATE Initialize $g(\tau^{(i)}, X_{\tau^{(i)}})$ as $g(T, X_T^{(i)})$ for all samples.
\FOR {$k = N-1:-1:1$}
    \STATE Find $i\in \{1,\dots, N_L\}$ such that $g(t_k, X_k^{(i)}) \ge \mathcal{C}_{k,\theta}(X_k^{(i)})$ and then update $g(\tau^{(i)}, X_{\tau^{(i)}})$ by $g(t_k, X_k^{(i)})$.
\ENDFOR
\STATE Calculate $\hat{L} = \frac{1}{N_L}\sum_{i=1}^{N_L} g(\tau^{(i)}, X_{\tau^{(i)}})$ and confidence interval.
\end{algorithmic}
\end{algorithm}

\subsection{Upper bound} \label{subsec:upper-bound}
The upper bound is derived from the dual formulation of optimal stopping problems, first introduced in \cite{rogers2002monte, haugh2004pricing}.  The optimal value $V_k = \sup_{\tau \in \mathcal{J}, \tau \ge t_k}\bE[g(\tau, X_\tau) | X_k]$, $k = 0,\dots, N$, is the smallest supermartingale that dominates the reward process $(g(t_k, X_k))_{k=0}^N$. By Doob-Meyer decomposition, $V_k$ can be decomposed into
\begin{equation} \label{eq:doob-meyer-decom}
    V_k = V_0 + M_k - A_k,
\end{equation}
where $(M_k)_{k=0}^N$ is a martingale and $(A_k)_{k=0}^N$ is a nondecreasing predictable process with $M_0 = A_0 = 0$. Now, let $(\Tilde{M}_k)_{k=0}^N$ be an arbitrary martingale with $\Tilde{M}_0 = 0$. For any stopping time $\tau \in \mathcal{J}$, we have
\begin{equation*}
    \bE[g(\tau, X_\tau)] = \bE[g(\tau, X_\tau) - \Tilde{M}_\tau] \le \bE\left[\max_{k=0,\dots,N}\left( g(t_k, X_k) - \Tilde{M}_k\right)\right].
\end{equation*}
Since this inequality holds for all stopping time, using any martingale $\Tilde{M}$, we can obtain an upper bound
\begin{equation} \label{eq:arbitrary-upper-bound}
    V_0 \le \bE\left[\max_{k=0,\dots,N}\left( g(t_k, X_k) - \Tilde{M}_k\right)\right].
\end{equation}
Moreover, by taking $\Tilde{M}_k = M_k$ with the Doob martingale $M$ in \eqref{eq:doob-meyer-decom}, we have
\begin{equation} \label{eq:doob-martingale-upper-bound}
    \bE\left[\max_{k=0,\dots,N}\left( g(t_k, X_k) - M_k\right)\right] = V_0 + \bE\left[\max_{k=0,\dots,N}\left( g(t_k, X_k) - V_k - A_k\right)\right].
\end{equation}
Since $g(t_k, X_k) - V_k \le 0$ and $A_k\ge 0$ for all $k=0,\dots,N$, the last expectation in \eqref{eq:doob-martingale-upper-bound} is nonpositive. By combining \eqref{eq:doob-martingale-upper-bound} and \eqref{eq:arbitrary-upper-bound}, we deduce that the equality in \eqref{eq:arbitrary-upper-bound} holds when using the Doob martingale $(M_k)_{k=0}^N$ in \eqref{eq:doob-meyer-decom}:
\begin{equation*}
    V_0 = \bE\left[\max_{k=0,\dots,N}\left( g(t_k, X_k) - M_k\right)\right].
\end{equation*}

Our estimate of the upper bound is based on \eqref{eq:arbitrary-upper-bound}. Since the upper bound is tight when using the Doob martingale, we aim to compute a martingale close to $(M_k)_{k=0}^N$ in \eqref{eq:doob-meyer-decom}. In fact, we have the following martingale representation, where the integrand includes the gradient $\nabla_x c(s, X_s)$ of continuation value $c(s,X_s)$. This type of martingale representation was used in \cite{belomestny2009true}, where they employed the identity $M_k = \int_0^{t_k} Z_t\,\dd W_t$ for a squared integrable process $(Z_t)_{0\le t\le T}$ and a regression method to compute $Z_t$.
\begin{prop} \label{prop:doob-martinagle-representation}
The Doob martingale $(M_k)_{k=0}^N$ in \eqref{eq:doob-meyer-decom} admits the martingale representation
\begin{equation*}
    M_k = \int_0^{t_k} \nabla_x c(s, X_s)^\top \sigma(X_s)\,\dd W_s,
\end{equation*}
where $c(t, X_t)$ is the continuation value defined in \eqref{eq:defi-cv-tau}.
\end{prop}

Recall that, at time step $t_j$, we approximate $\nabla_x c(t_j, X_j)$ by a neural network $\mathcal{G}_{j,\theta}$. The martingale part $\Delta M_j$ in the proposed loss function \eqref{eq:novel-loss} leads to a natural approximation of the Doob martingale
\begin{equation} \label{eq:direct-doob-martingale-approx}
    M_k = \sum_{j=0}^{k-1}\int_{t_j}^{t_{j+1}} \nabla_x c(s, X_s)^\top \sigma(X_s)\,\dd W_s \approx M_1 - M_0 + \sum_{j=1}^{k-1} \Delta M_j,
\end{equation}
with $\Delta M_j := \mathcal{G}_{j,\theta}(X_j)^\top \sigma(X_j)\Delta W_j$.

Next, we present a simple approach to further improve the accuracy of Doob martingale approximation. One can observe from \eqref{eq:direct-doob-martingale-approx} that the direct approximation using $\Delta M_j$ is subject to the time discretization level $\Delta t$. To improve the accuracy of martingale approximation, we can use finer time grid to approximate $M_{k+1} - M_{k} = \int_{t_{k}}^{t_{k+1}} \nabla_x c(s, X_s)^\top \sigma(X_s)\,\dd W_s$. Over $[t_{k}, t_{k+1}]$ for $k=1,\dots, N-1$, we consider a finer time grid $t_{k} = t_{k,0} < t_{k,1} < \dots < t_{k,J} = t_{k+1}$ with $\delta t = \Delta t/J$, $t_{k,j} = t_{k} + j\delta t$ for $j=0,\dots, J$. Then, using $\nabla_x c(s, x) \approx \nabla \hat{c}_{k}(x) \approx \mathcal{G}_{k,\theta}(x)$ for $s\in [t_{k}, t_{k+1}]$, cf. \eqref{eq:cv-linear-bsde}-\eqref{eq:noval-loss-init}, we obtain the approximation
\begin{equation} \label{eq:martingale-difference-approx}
    M_{k+1} - M_{k} = \sum_{j=0}^{J-1} \int_{t_{k,j}}^{t_{k,j+1}} \nabla_x c(s, X_s)^\top \sigma(X_s)\,\dd W_s \approx \sum_{j=0}^{J-1} \mathcal{G}_{k,\theta}(X_{t_{k,j}})^\top \sigma(X_{t_{k,j}})\left(W_{t_{k,j+1}} - W_{t_{k,j}}\right).
\end{equation}

To obtain an upper bound, it remains to approximate $M_1 - M_0$. Recall that $M_0 = 0$ and $M_1 = V_1 - \bE[V_1|X_0]$. Using the trained neural network $\mathcal{C}_{1,\theta}$ and $N_U$ samples $X_1^{(i)}$ of $X_1$ for $i=1,\dots, N_U$, we approximate the martingale difference of the $i$-th sample by
\begin{equation} \label{eq:martingale-difference-approx-1st}
    M_1^{(i)} - M_0^{(i)} = \max\left( g(t_1, X_1^{(i)}), \mathcal{C}_{1,\theta}(X_1^{(i)}) \right) - \frac{1}{N_U}\sum_{i=1}^{N_U} \max\left( g(t_1, X_1^{(i)}), \mathcal{C}_{1,\theta}(X_1^{(i)}) \right).
\end{equation}
Finally, the methodology for the true upper bound is summarized in Algorithm \ref{alg:upper-deepBSDEOS}.

\begin{algorithm}[htbp!]
\caption{Computing upper bound}
\label{alg:upper-deepBSDEOS}
\begin{algorithmic}[1]
\REQUIRE Trained networks $\mathcal{C}_{1,\theta}$ and $\mathcal{G}_{k,\theta}$, $k=1,\dots, N-1$. Sample size $N_U$. Finer time grid level $J$.
\ENSURE Upper bound estimate $\hat{U}$, confidence interval
\STATE Generate $N_U$ samples $(X_{k,j}^{(i)})_{k=0:N, j=0:J}$, $i=1,\dots, N_U$, on finer time grid of forward SDEs \eqref{eq:Ito-diffusion}.
\STATE Compute $M_{k+1}^{(i)} - M_k^{(i)}$ for $k=1,\dots,N-1$, $i=1,\dots, N_U$, using the approximation \eqref{eq:martingale-difference-approx}.
\STATE Compute $M_1^{(i)} - M_0^{(i)}$ using the approximation \eqref{eq:martingale-difference-approx-1st}.
\STATE Calculate $M_k^{(i)}$ from steps 2 and 3.
\STATE Calculate $\hat{U}=\frac{1}{N_U}\sum_{i=1}^{N_U} \max_{k=0,\dots,N} \left(g(t_k, X_k^{(i)}) - M_k^{(i)}\right)$ and confidence interval.
\end{algorithmic}
\end{algorithm}

\section{Bias analysis}\label{sec:stop-value}
We analyze the approximation bias of the proposed scheme using stopping time iteration scheme \eqref{eq:dp-stopping-time}. We argue that its low-biased property makes the stopping time iteration more stable than the high-biased value iteration scheme \eqref{eq:dp-value-iteration}.
The proposed deep primal-dual BSDE framework based on \eqref{eq:noval-loss-init} is closely related to the reflected deep backward dynamic programming (RDBDP) scheme proposed in \cite{hure2020deep} and its variant in \cite{chen2021deep}. The RDBDP scheme was developed using reflected BSDE, which can be viewed as a value iteration approach.

To focus on the bias analysis of stopping time iteration and value iteration, we make the following assumption which asserts that the conditional expectation can be approximated unbiasedly.
\begin{assu} \label{assum:unbiased-approx}
There exists an algorithm that can compute an unbiased estimate $\hat{f}_k$ of $f_k(x) = \bE[\xi|X_k = x]$ for any random variable $\xi$ and $k=1,2,\dots, N-1$. We denote this unbiased algorithm by $\mathcal{E}_k$ with $\hat{f}_k(X_k) = \mathcal{E}_k(\xi)$.
\end{assu}
Under Assumption \ref{assum:unbiased-approx}, the approximation based on the value iteration \eqref{eq:dp-value-iteration} is biased high, which has been mentioned in \cite{carriere1996valuation, stentoft2014value}. We present the following variant of the high-biased property in Proposition \ref{prop:value-high-bias}, and the low-biased property of stopping time iteration \eqref{eq:dp-stopping-time} in Proposition \ref{prop:stopping-time-low-bias}. The proofs of the propositions are in Appendix \ref{append:proofs}.

\begin{prop} \label{prop:value-high-bias}
Under Assumption \ref{assum:unbiased-approx}, given the exact value $V_{k+2}$ at time $t_{k+2}$ for any fixed $k\in \{1,2, \dots, N-2\}$, the true continuation value at time $t_k$ based on the value iteration \eqref{eq:dp-value-iteration}, i.e.,
$$C_k:= \bE[\max\left( g(t_{k+1}, X_{k+1}), \bE[V_{k+2}|X_{k+1}]\right) | X_k],$$
is approximated using the unbiased algorithm $\mathcal{E}_{k+1}$ and $\mathcal{E}_k$ by
\begin{align*}
\hat{C}_{k+1} &= \mathcal{E}_{k+1}(V_{k+2}), \\
\hat{V}_{k+1} &= \max\left( g(t_{k+1}, X_{k+1}), \hat{C}_{k+1}\right), \\
\hat{C}_k &= \mathcal{E}_k(\hat{V}_{k+1}).
\end{align*}
Then, the estimate $\hat{C}_k$ is biased high for $C_k$.
\end{prop}

\begin{prop} \label{prop:stopping-time-low-bias}
Under Assumption \ref{assum:unbiased-approx}, given the exact optimal stopping time $\tau_{k+2}$ at time $t_{k+2}$ for any fixed $k\in \{1,2, \dots, N-2\}$, the true continuation value at time $t_k$ based on the stopping time iteration \eqref{eq:dp-stopping-time}, i.e.,
\begin{equation} \label{eq:exact-Ck-stopping-time}
    C_k = \bE[g(\tau_{k+1}, X_{\tau_{k+1}}) | X_k]\quad\text{with } \tau_{k+1} = \begin{cases}
        t_{k+1}, &\text{ if } g(t_{k+1}, X_{k+1}) \ge \bE[g(\tau_{k+2}, X_{\tau_{k+2}}) | X_{k+1}] \\
        \tau_{k+2}, &\text{ if } g(t_{k+1}, X_{k+1}) < \bE[g(\tau_{k+2}, X_{\tau_{k+2}}) | X_{k+1}]
    \end{cases},
\end{equation}
is approximated using the unbiased algorithm $\mathcal{E}_{k+1}$ and $\mathcal{E}_k$ by
\begin{align*}
\tilde{C}_{k+1} &= \mathcal{E}_{k+1}(g(\tau_{k+2}, X_{\tau_{k+2}})), \\
\tilde{\tau}_{k+1} &= \begin{cases}
        t_{k+1}, \quad &\text{ if } g(t_{k+1}, X_{k+1}) \ge \tilde{C}_{k+1} \\
        \tau_{k+2}, \quad &\text{ if } g(t_{k+1}, X_{k+1}) < \tilde{C}_{k+1}
    \end{cases} \\
\tilde{C}_k &= \mathcal{E}_k(g(\tilde{\tau}_{k+1}, X_{\tilde{\tau}_{k+1}})).
\end{align*}
Then, the estimate $\tilde{C}_k$ is biased low for $C_k$.
\end{prop}

Next, we show that the RDBDP scheme for the optimal stopping problem results in an approximation scheme for the value iteration \eqref{eq:dp-value-iteration}, and thus is biased high by Proposition \ref{prop:value-high-bias}. For the optimal stopping problem \eqref{eq:os-original-continuous}, the optimal value $Y_t$ satisfies the following reflected BSDE:
\begin{equation} \label{eq:RBSDE}
    \begin{cases}
        Y_t &= g(T, X_T) - \int_t^T Z_s^\top \,\dd W_s + A_T - A_t \\
        Y_t &\ge g(t, X_t), \quad \int_0^T \left(Y_t - g(t, X_t)\right) \,\dd A_t = 0,
    \end{cases}
\end{equation}
where $(A_t)_{0\le t\le T}$ is a non-decreasing process corresponding to the continuous version of $(A_k)_{k=0}^N$ in \eqref{eq:doob-meyer-decom}. The RDBDP scheme \cite[Section 3.3]{hure2020deep} employs neural networks $(\mathcal{U}_{k,\theta}(\cdot), \mathcal{Z}_{k,\theta}(\cdot))$ to approximate the reflected BSDE solution $(Y_{t_k}, Z_{t_k})$ for $k = 0,\dots, N-1$. It updates the value backwards\footnote{We drop the generator term $f$ in RDBDP scheme corresponding to \eqref{eq:RBSDE}.} by first setting $\hat{\mathcal{U}}_{N} = g(T, \cdot)$, and then solving for $k = N-1, \dots, 0$,
\begin{equation*}
    \begin{aligned}
        \theta^*_k &= \argmin_{\theta} \bE\left[\big|\hat{\mathcal{U}}_{k+1}(X_{k+1}) - \mathcal{U}_{k,\theta}(X_k) - \mathcal{Z}_{k,\theta}(X_k)^\top \Delta W_k \big|^2\right], \\
        \hat{\mathcal{U}}_{k} &= \max\left( g(t_k, \cdot), \mathcal{U}_{k,\theta_k^*} \right).
    \end{aligned}
\end{equation*}
One can view $\mathcal{U}_{k,\theta_k^*}(X_k)$ as a neural network approximation for $\mathbb{E}[V_{k+1}|X_k]$ and $\hat{\mathcal{U}}_{k}(X_{k})$ for $V_{k}$ in the value iteration \eqref{eq:dp-value-iteration}. Consequently, if we assume that training neural networks leads to an unbiased estimate for $\mathbb{E}[V_{k+1}|X_k]$, then the RBSDE scheme is still biased high.

\begin{figure}[htbp!]
    \centering
    \includegraphics[width=0.7\linewidth]{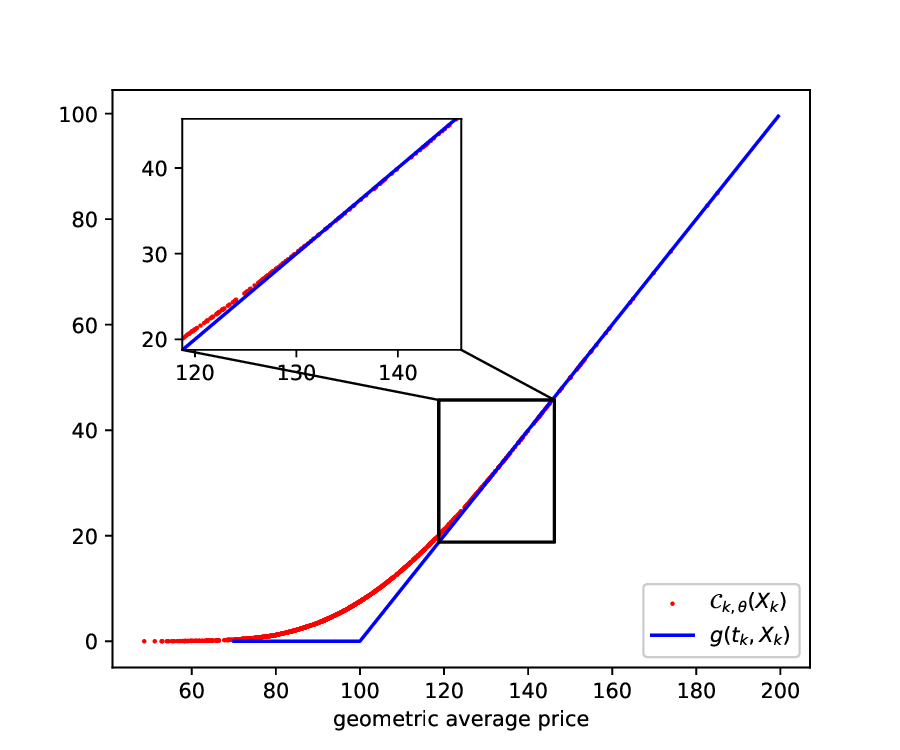}
    \caption{The approximative continuation value $\mathcal{C}_{k,\theta}(X_k)$ after training and the immediate stopping reward $g(t_k, X_k)$, where the $x$-axis represents the geometric average of $d=20$ components of $X_k$. $k=25$, $N=50$, $\Delta t = T/N=0.04$.}
    \label{fig:bias_geobaskcall_20}
\end{figure}

The systematic high-biased property poses a substantial challenge on comparing the immediate stopping reward $g(t_k, X_k)$ with the approximate continuation value $\mathcal{U}_{k,\theta_k^*}(X_k)$. When $\Delta t$ approaches zero, the difference between these two functions is small in the stopping region. We illustrate this with pricing a $20$-dimensional geometric basket call option, see Section \ref{subsec:ex1-geobaskcall} for details. Figure \ref{fig:bias_geobaskcall_20} depicts the approximate continuation value $\mathcal{C}_{k,\theta}(X_k)$ and $g(t_k, X_k)$ for $t_k=1$ with $T = 2$ using the trained network with Algorithm \ref{alg:deepBSDEOS}. One observes from the figure that in the stopping region, i.e., $g(t_k, X_k)\ge \mathcal{C}_{k,\theta}(X_k)$, two functions are close to each other. Then if the approximation for continuation value is biased high, computing the correct stopping region will be challenging. A remedy for this issue is to use mixed-type values for training from practical perspective \cite[Section 5.2]{chen2021deep}. Our method based on stopping time iteration is naturally biased low, cf. Proposition \ref{prop:stopping-time-low-bias}. Thus, it is more stable to compute the stopping region than the high-biased value iteration scheme \eqref{eq:dp-value-iteration}.

\section{Experiments}\label{sec:num}
In this section, we test the efficacy of the proposed method in American or Bermudan option pricing problems. Pricing an American option leads to solving a continuous-time optimal stopping problem, while the Bermudan option amounts to the discrete-time counterpart. For all examples, FNNs with $2$ hidden layers are employed for the subnetworks $\mathcal{C}_{k,\theta}$ and $\mathcal{G}_{k,\theta}$ for all $k=1,\dots, N-1$. We focus on the numerical results in the main text and defer technical details and hyperparameter settings to Appendix \ref{append:hyperparameters}. 
The code for experiments can be found in the public GitHub repository \url{https://github.com/jiefeiy/deep-primal-dual-BSDE}.

\subsection{American geometric basket call option} \label{subsec:ex1-geobaskcall}
The first example is pricing American geometric basket call options with $d$ underlying assets, which was considered in \cite[Section 4.3]{sirignano2018dgm}, \cite[Section 8.1]{chen2021deep}, and \cite[Section 6.2]{yang2024gradient}. It serves as the benchmark of the high-dimensional optimal stopping problem since analytically it can be reduced to a one-dimensional problem. We compute the reference exact value of the reduced problem by the Crank-Nicolson discretization with the penalty method using 15000 spatial grid points, 3000 time steps, and truncated spatial domain $[0, 4\overline{X}_0]$, $\overline{X}_0 = (\prod_{j=1}^d X_{0,j})^{1/d}$. The dynamic of forward process $(X_t)_{0\le t\le T}$ follows the multi-asset Black-Scholes model, i.e., \eqref{eq:Ito-diffusion} with $\mu(X_t) = (r - \delta)X_t, r=0, \delta=0.02$, $\sigma(X_t) = 0.25\op{diag}(X_t)$, and $\bE[\dd W_{t,i}\dd W_{t,j}] = \rho_{ij}\dd t, \rho_{ij}=0.75, i\ne j$. The time horizon is $T=2$ and we take an equidistant time grid with $\Delta t = T/N$, $N=50$. The reward function $g$ represents the discounted payoff of the geometric basket call option, which is $g(t, x) = e^{-rt}\max( (\prod_{j=1}^d x_j)^{1/d} - K, 0)$ with strike price $K = 100$.

We present in Table \ref{tab:geobask call deep bsde} the training time\footnote{The experiments were performed with PyTorch 2.2.2 on an Intel Core i9-10900 CPU 2.8 GHz desktop with 64GB DDR4 memory. The training time is expected to be further reduced with a GPU system.} in Algorithm \ref{alg:deepBSDEOS}, lower and upper bounds, and their $95\%$ confidence intervals by Algorithms \ref{alg:lower-deepBSDEOS} and \ref{alg:upper-deepBSDEOS}.  The training time is larger for higher dimensional problems, since we use wider hidden layers. The training time for $d=200$ is smaller than $d=100$, since we use half of the batch size for $d=200$ due to the memory limitation of the computing device. To show that the training process is stable for different random seeds, we report the results of the $100$-dimensional pricing problem for $5$ independent repeated runs in Table \ref{tab:repeat-geobaskcall-d100}.

\begin{table}[htbp!]
    \centering
    \begin{tabular}{ccccccc}
        \hline
        $d$ & $x_{0,j}$ & exact & training time & lower bound & upper bound \\
        \hline
        3 & 100 & 10.7192 & 384s & $10.7111 \,(\pm 0.0211)$ & $10.7984 \,(\pm 0.0047)$ \\
        20 & 100 & 10.0333 & 515s & $10.0057 \,(\pm 0.0195)$ & $10.1341 \,(\pm 0.0050)$ \\
        100 & 100 & 9.9352 & 783s & $9.9058 \,(\pm 0.0187)$ & $10.1365 \,(\pm 0.0073)$\\
        200 & 100 & 9.9229 & 706s & $9.9037 \,(\pm 0.0270)$ & $10.1556 \,(\pm 0.0092)$ \\
        \hline
    \end{tabular}
    \caption{Results for pricing American geometric basket call options with $d$ assets: $N_L = 2^{21}$, $N_U = 2^{15}$, and $J=32$.}
    \label{tab:geobask call deep bsde}
\end{table}

\begin{table}[hbt!]
    \centering
    \begin{tabular}{ccc}
        \hline
         & lower bound & upper bound \\
        \hline
        test 1 & $9.9012 \,(\pm 0.0184)$ & $10.1565 \,(\pm 0.0074)$ \\
        test 2 & $9.9131 \,(\pm 0.0191)$ & $10.1281 \,(\pm 0.0071)$ \\
        test 3 & $9.9187 \,(\pm 0.0195)$ & $10.1575 \,(\pm 0.0073)$ \\
        test 4 & $9.9180 \,(\pm 0.0197)$ & $10.1255 \,(\pm 0.0074)$ \\
        test 5 & $9.9084 \,(\pm 0.0197)$ & $10.1038 \,(\pm 0.0071)$ \\
        \hline
    \end{tabular}
    \caption{Repeated tests for $d=100$. Parameters are the same as those in Table \ref{tab:geobask call deep bsde}.}
    \label{tab:repeat-geobaskcall-d100}
\end{table}

\begin{figure}[htbp!]
    \centering
    \subfloat[$d=20$\label{fig:geobaskcall_20}]{\includegraphics[width = .45\textwidth]{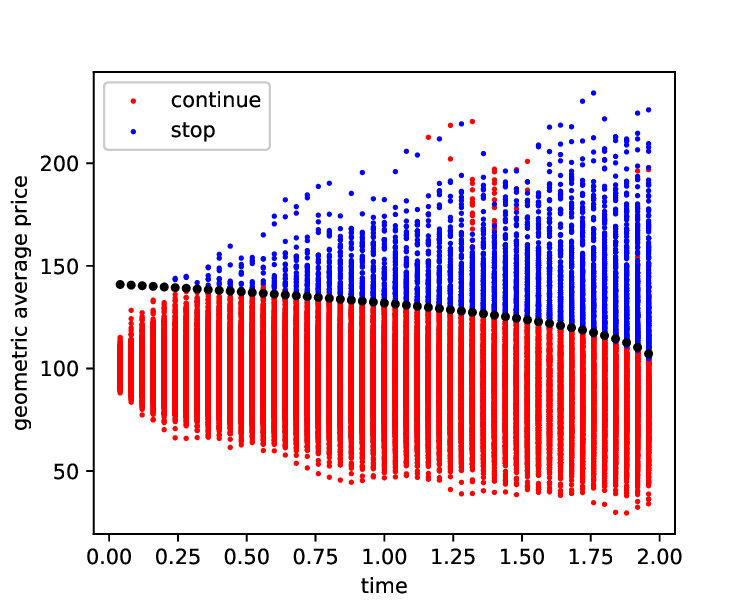}}
    \subfloat[$d=100$\label{fig:geobaskcall_100}]{\includegraphics[width = .45\textwidth]{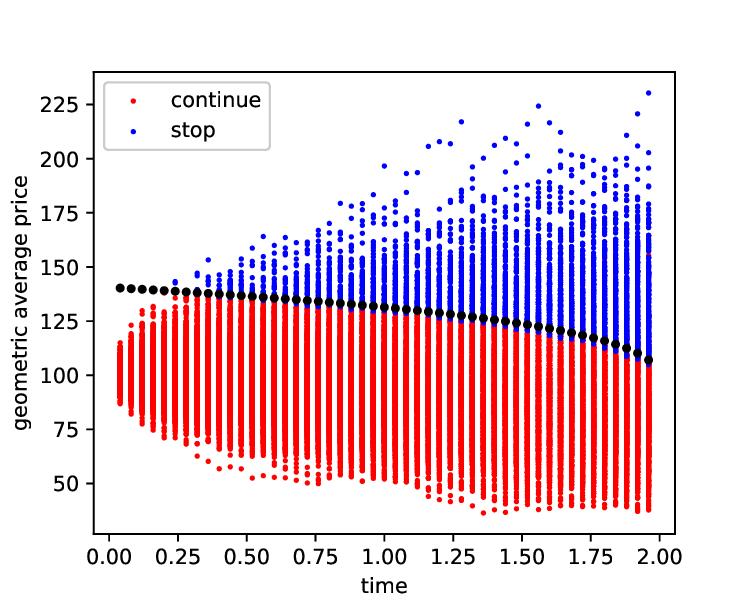}}
    \caption{Classification of the simulated continued and stopped data using the trained networks for $d$-dimensional problems. Black dots represent the exact optimal stopping boundary.}
    \label{fig:geobask_call_labels_d20_d100}
\end{figure}

To visualize the accuracy of the proposed method for optimal stopping time, we illustrate the stopping and continuation region in Figure \ref{fig:geobask_call_labels_d20_d100} with respect to the geometric average of $d$ components of $X_k$ for $k=1,\dots, N-1$. We sample $1000$ paths of $(X_k)_{k=1}^{N-1}$ and classify the sample points using the trained networks $\mathcal{C}_{k,\theta}$. Compared with the exact optimal stopping boundary (black dots in Figure \ref{fig:geobask_call_labels_d20_d100}), our method can effectively split the spatial domain into stopping and continuation region.

Next, we display the results for hedging ratio. The hedging strategy can be constructed with the so-called delta, i.e., the first order derivative of the value function. Since pricing a geometric basket option can be reduced to a one-dimensional problem, we illustrate delta values by considering $\frac{\partial V_k}{\partial \overline{X}_k}$, where $\overline{X}_k := \left(\prod_{j=1}^d X_{k,j}\right)^{1/d}$. Using $V_k = \max\left(g(t_k, X_k), c(t_k, X_k)\right)$, we have $\frac{\partial V_k}{\partial \overline{X}_k} = \frac{\partial c(t_k, X_k)}{\partial \overline{X}_k}$ in the continuation region and $\frac{\partial V_k}{\partial \overline{X}_k} = \frac{\partial g(t_k, X_k)}{\partial \overline{X}_k}$ in the stopping region. The trained gradient subnetwork $\mathcal{G}_{k,\theta}$ approximates $\nabla_x c(t_k, \cdot)$ leading to the evaluation of delta values directly. Figure \ref{fig:delta_geobaskcall} shows the projected delta values using $\mathcal{G}_{k,\theta}$, see Appendix \ref{append:hyperparameters} for details. The exact delta is computed from the reduced one-dimensional problem by quadrature and interpolation method for the Bermudan option with $N=50$ exercise opportunities. Although the smooth fit condition indicates the value function of an American option has $C^1$ regularity, the value of a Bermudan option, defined by the maximum of payoff and continuation value, can involve a kink (discontinuity of the first order derivative) in the spatial domain. Thus, one observes the discontinuity in Figure \ref{fig:delta_geobaskcall}. The comparison of the projected delta values using the trained networks with the reference value shows that our method can approximate the delta efficiently. Note that the delta function in a $d$-dimensional problem can be viewed as a map from $\R^d$ to $\R^d$, which is difficult to compute using classical grid-based approaches. The efficiency of the deep primal-dual BSDE method demonstrates the capability of neural networks for high-dimensional approximation.

\begin{figure}[htbp!]
    \centering
    \subfloat[$d=20$\label{fig:delta_geobaskcall_20}]{\includegraphics[width = .45\textwidth]{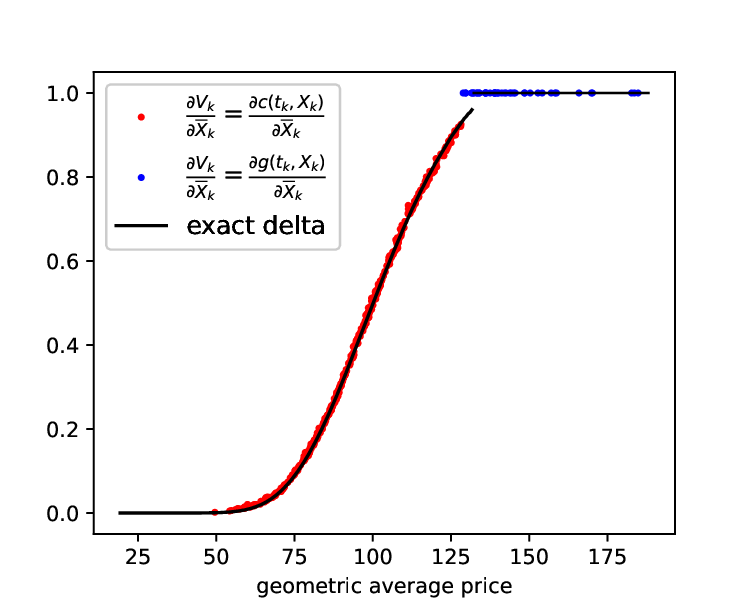}}
    \subfloat[$d=100$\label{fig:delta_geobaskcall_100}]{\includegraphics[width = .45\textwidth]{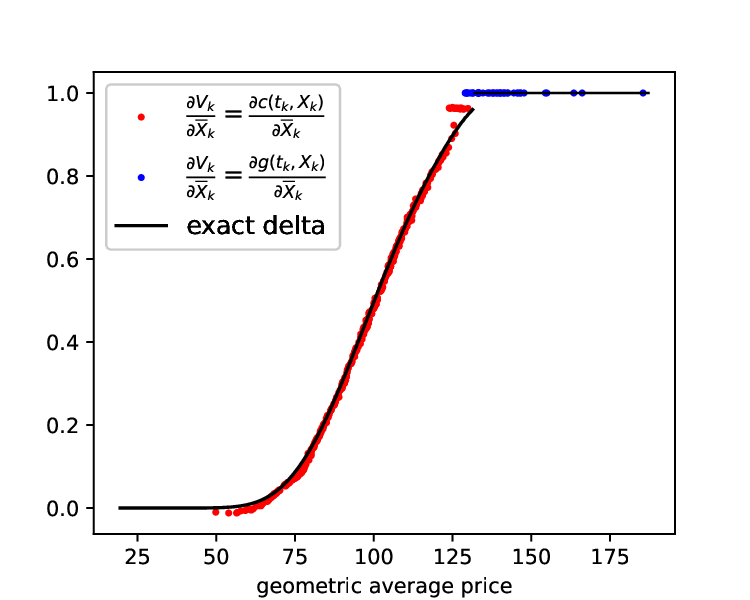}}
    \caption{Projected delta values at $t_k=1$ of $500$ simulated data points using the trained networks for $d$-dimensional problems. }
    \label{fig:delta_geobaskcall}
\end{figure}

\subsection{American strangle spread basket option} \label{subsec:ex2-strangle}
Next, we test the proposed method for pricing an American strangle spread basket option depending on five correlated stocks, see \cite[Section 4]{kohler2010pricing} and \cite[Section 4.4.2]{becker2021solving}. The dynamic of five correlated stock price $(X_t)_{0\le t\le T}$, is given by the diffusion process in \eqref{eq:Ito-diffusion} with drift and volatility being
\begin{equation*}
    \mu(X_t) = rX_t, r=0.05, \quad\text{and}\quad \sigma(X_t) = \op{diag}(X_t) \begin{bmatrix}
        0.3024 &  0.1354 &  0.0722  & 0.1367 &  0.1641 \\
    0.1354 &  0.2270 &  0.0613 &  0.1264 &  0.1610 \\
    0.0722 &  0.0613 &  0.0717 &  0.0884 &  0.0699 \\
    0.1367 &  0.1264 &  0.0884 &  0.2937 &  0.1394 \\
    0.1641 &  0.1610 &  0.0699 &  0.1394 &  0.2535
    \end{bmatrix}.
\end{equation*}
The components of $d$-dimensional driven Brownian motion $W_t$ are assumed to be independent. The option has expiration time $T=1$ and we take $N=48$ equidistant time points.
The discounted payoff of the strangle spread basket option is given by
\begin{equation*}
    g(t, X_t) = e^{-rt} \max\Big( 15 - \max\Big( 25 - \Big|\frac{1}{d}\sum_{j=1}^d X_{t,j} - 100 \Big|, 0\Big), 0\Big).
\end{equation*}

Table \ref{tab:strangle-bounds} shows the lower and upper bounds by our method and the reference price in \cite{becker2021solving, kohler2010pricing}. We note that the two reference methods compute suboptimal stopping strategy leading to lower bounds of the original problem. The lower bound by the proposed approach is similar to the price given by \cite{becker2021solving}, while higher than that of \cite{kohler2010pricing}. This indicates an improvement of accuracy of the deep primal-dual BSDE method. The small difference between lower and upper bounds indicates reliability of our method for the five-dimensional optimal stopping problem.

\begin{table}[htbp!]
    \centering
    \begin{tabular}{ccccc}
        \hline
        $d$ & lower bound & upper bound & \cite{becker2021solving} & \cite{kohler2010pricing}\\
        \hline
        5 & $11.7981 \,(\pm 0.0068)$ & $11.8665 \,(\pm 0.0056)$ & 11.797 & 11.75 \\
        \hline
    \end{tabular}
    \caption{Bounds with 95\% confidence intervals for an American strangle spread basket option with $48$ exercise opportunities: $N_L = 2^{21}$, $N_U = 2^{15}$, and $J = 32$.}
    \label{tab:strangle-bounds}
\end{table}

\subsection{Bermudan put option under Heston model} \label{subsec:ex3-put-heston}
The Heston model defines the dynamic of log-price process, $X_{t,1}$, and volatility process, $X_{t,2}$, by two-dimensional stochastic differential equations:
\begin{align*}
    \dd X_{t,1} &= (r-\tfrac{1}{2}X_{t,2})\,\dd t + \sqrt{X_{t,2}} \left(\rho \,\dd W_{t,1} + \sqrt{1-\rho^2}\,\dd W_{t,2} \right), \quad X_{0,1} = 0, \\
    \dd X_{t,2} &= \kappa(\theta - X_{t,2})\,\dd t + \nu \sqrt{X_{t,2}}\,\dd W_{t,1}, \quad X_{0,2} = v_0,
\end{align*}
where $W_{t,1}$ and $W_{t,2}$ are two independent Brownian motions, and the model parameters $r$, $\kappa$, $\theta$, $\nu$, $\rho$, and $v_0$ represent the interest rate, the speed of mean reversion, the mean level of variance, the variance of volatility process, the correlation coefficient, and the initial volatility, respectively. The discounted payoff $g$ of the put option is given by $g(t, X_t) = e^{-rt}\max\left(K- s_0\exp(X_{t,1}), 0\right)$. We consider the Bermudan put with $N=50$ equidistant exercises opportunities.

Table \ref{tab:heston-bounds} shows the lower and upper bounds computed by the proposed deep primal-dual BSDE method and the reference price by the COS method \cite{fang2011fourier} for different initial underlying asset price $s_0$. We note that the grid-based COS method is expected to be more accurate than the deep learning-based method in this two-dimensional example. The proposed method can be extended easily to multi-factor or multi-asset stochastic volatility models in high dimensions, while the conventional grid-based methods usually become infeasible due to the curse of dimensionality.

\begin{table}[htbp!]
    \centering
    \begin{tabular}{cccc}
        \hline
        $s_0$ & lower bound & upper bound & COS \cite{fang2011fourier} \\
        \hline
        9 & $1.1058 \,(\pm 0.0007)$ & $1.1084 \,(\pm 0.0002)$ & 1.1061 \\
        10 & $0.5181 \,(\pm 0.0006)$ & $0.5230 \,(\pm 0.0002)$ & 0.5186 \\
        11 & $0.2123 \,(\pm 0.0004)$ & $0.2177 \,(\pm 0.0002)$ & 0.2131 \\
        \hline
    \end{tabular}
    \caption{Bounds with 95\% confidence intervals for Bermudan put options under Heston model with $50$ exercise opportunities: $N_L = 2^{22}$, $N_U = 2^{15}$, $J = 32$, $r=0.1$, $\kappa=5$, $\theta=0.16$, $\nu=0.9$, $\rho=0.1$, $v_0=0.0625$, $T=0.25$, and $K=10$.}
    \label{tab:heston-bounds}
\end{table}

\subsection{Bermudan max-call option} \label{subsec:ex4-max-call}
The last example is pricing Bermudan max-call options. While the Bermudan max-call options have been widely studied in literature, see \cite{glasserman2004monte, haugh2004pricing, andersen2004primal, belomestny2009true, bayer2023pricing, becker2019deep}, we consider a more challenging case with $N=100$ exercise opportunities compared with the classical case of $9$ exercise opportunities among a three-year period. Such large exercise opportunities can also be used to approximate the American option price, which has also been considered in \cite[Section 8.2]{chen2021deep}. The dynamic of forward process $(X_t)_{0\le t\le T}$ is given by the diffusion process in \eqref{eq:Ito-diffusion} with $\mu(X_t) = (r-\delta)X_t$, $r=0.05, \delta = 0.1$, and $\sigma(X_t) = 0.2\op{diag}(X_t)$. Here, the components of d-dimensional driven Brownian motion $W_t$ are assumed to be independent. The option has expiration time $T=3$ and can be early exercised at $N=100$ equidistant time points. The reward function $g$ is $g(t, x) = e^{-rt} \max\left( \max_{1\le j\le d}x_j - K, 0\right)$ with strike price $K=100$.

Table \ref{tab:maxcall-bounds} shows the results for $d$-dimensional problems with different initial values of the forward process. In the case of $d=5$, we quote the price from \cite{chen2021deep} for reference. The reference approach also uses a suboptimal stopping strategy to derive the final result, and thus it yields a lower bound for the true value. The deep primal-dual BSDE method gives slightly higher lower bounds in all cases of $d=5$, and therefore, provides improved accuracy.

\begin{table}[htbp!]
    \centering
    \begin{tabular}{cccccc}
        \hline
        $d$ & $x_{0,j}$ & lower bound & upper bound & \cite{chen2021deep} \\
        \hline
        2 & 90 & $8.2420 \,(\pm 0.0151)$ & $8.3420 \,(\pm 0.0029)$ \\
        2 & 100 & $14.1969 \,(\pm 0.0188)$ & $14.2729 \,(\pm 0.0031)$ \\
        2 & 110 & $21.7661 \,(\pm 0.0225)$ & $21.8086 \,(\pm 0.0034)$ \\
        5 & 90 & $16.9405 \,(\pm 0.0212)$ & $17.2241 \,(\pm 0.0076)$ & 16.8896 \\
        5 & 100 & $26.5927 \,(\pm 0.0248)$ & $26.8561 \,(\pm 0.0083)$ & 26.4876 \\
        5 & 110 & $37.2724 \,(\pm 0.0288)$ & $37.5932 \,(\pm 0.0092)$ & 37.0996 \\
        10 & 90 & $26.5888 \,(\pm 0.0242)$ & $27.5517 \,(\pm 0.0215)$ \\
        10 & 100 & $38.7794 \,(\pm 0.0278)$ & $39.9960 \,(\pm 0.0262)$ \\
        10 & 110 & $51.3909 \,(\pm 0.0310)$ & $52.3989 \,(\pm 0.0242)$ \\
        \hline
    \end{tabular}
    \caption{Bounds with 95\% confidence intervals for Bermudan max-call options with $100$ exercise opportunities: $N_L = 2^{21}$, $N_U = 2^{15}$, and $J = 32$.}
    \label{tab:maxcall-bounds}
\end{table}

\section{Conclusion}\label{sec:conclusion}
This paper presents a novel deep learning-based approach for solving optimal stopping problems. It is scalable in high dimensions and reliable by computing lower and upper bounds. The upper bound is computed in a non-nested Monte Carlo method, which improves the efficiency compared with the nested simulation approach, especially when the number of time steps is large. Using the stopping time iteration, the approximate continuation value is biased low for the truth. The low-biased property makes the scheme more stable than the high-biased value iteration. Moreover, we show that the proposed loss function has a variance-reduced stochastic gradient estimate compared with the neural network variants of the Longstaff-Schwartz method. Such variance reduction is crucial for the convergence of stochastic gradient descent algorithms. Numerical experiments up to $200$ dimensions demonstrate the efficacy of the proposed method.

There are several intriguing directions for future works. From practitioners' perspective, we only use classical FNNs with 2 hidden layers for the value and gradient subnetworks. It is promising to investigate whether other types of neural networks can further improve the accuracy. The proposed method is also applicable to various optimal stopping problems. There remains much work in applications of more challenging examples, such as more sophisticated dynamics of $X_t$ and discontinuous reward functions. From theoretical perspective, the proposed loss function will decay close to zero (subject to time discretization error) as the neural networks approach the targets. Inspired by the convergence analysis in \cite{han2020convergence} for the seminal deep BSDE method in \cite{E2017deepbsde}, we believe that the value of the proposed loss function may serve as an error indicator of the neural network approximation, and thus, will provide a criterion for determining when to stop the training.

\section*{Acknowledgments}
This work was supported by the University of Hong Kong under the HKU Presidential PhD Scholar Programme (HKU-PS); RGC under Early Career Scheme (Project number: 27301921).

\bibliographystyle{abbrv}
\bibliography{references}

\newpage
\appendix
\section{Proofs} \label{append:proofs}
\subsection{Proof of Proposition \ref{prop:var-stochastic-gradient}}
\begin{proof}
The first moments equal due to the zero mean of the Brownian motion increment $\Delta W_k$. Next, A combination of \eqref{eq:stochastic-grad-estimate-ls} and \eqref{eq:cv-linear-bsde} directly leads to 
\begin{align*}
    \bE\left[(\hat{G}^{\rm ls})^2 \right] = 4\bE[\partial_\theta \mathcal{C}_{k,\theta}(X_k)^2] \bE\left[ \left( \int_t^{\tau_{k+1}} \nabla_x c(s, X_s)^\top \sigma(X_s) \, \dd W_s \right)^2 \right].
\end{align*}
Then an application of the It\^{o} isometry yields \eqref{eq:variance-sg-ls}. To derive the last assertion, we first obtain from \eqref{eq:stochastic-grad-estimate-bsde} and \eqref{eq:cv-linear-bsde},
\begin{align*}
    \bE\left[ (\hat{G}^{\rm bsde})^2 \right] &= C \bE\left[ \left( \int_t^{\tau_{k+1}} \nabla_x c(s, X_s)^\top \sigma(X_s) \, \dd W_s - \sum_{j=k}^{\tau_{k+1}/\Delta t - 1} \mathcal{G}_{j,\theta}(X_j)^\top \sigma(X_j)\Delta W_j \right)^2 \right] \\
    &= C\bE\left[ \left(\sum_{j=k}^{\tau_{k+1}/\Delta t - 1} \int_{t_j}^{t_{j+1}} \left(\nabla_x c(s, X_s)^\top \sigma(X_s) - \mathcal{G}_{j,\theta}(X_j)^\top \sigma(X_j) \right) \,\dd W_s \right)^2 \right] \\
    &= C \sum_{j=k}^{\tau_{k+1}/\Delta t - 1} \bE\left[ \int_{t_j}^{t_{j+1}} \left(\nabla_x c(s, X_s)^\top \sigma(X_s) - \mathcal{G}_{j,\theta}(X_j)^\top \sigma(X_j) \right)^2 \,\dd s \right].
\end{align*}
Here, the last equality holds due to the independent increments of Brownian motion and the It\^{o} isometry. Finally, since $\Delta t = T/N$ and $\tau_{k+1}/\Delta t \le N$, we obtain
\begin{equation*}
    \bE\left[ (\hat{G}^{\rm bsde})^2 \right] \le CN \Delta t \sup_{t_j\le s\le t_{j+1}, j\ge k} \bE\left[ \left|\nabla_x c(s, X_s)^\top \sigma(X_s) - \mathcal{G}_{j,\theta}(X_j)^\top \sigma(X_j)\right|^2 \right],
\end{equation*}
which leads to the estimate \eqref{eq:variance-sg-bsde}.
\end{proof}

\subsection{Proof of Proposition \ref{prop:doob-martinagle-representation}}
\begin{proof}
For any $k = 0, \dots, N-1$, the subtraction of the Doob-Meyer decomposition \eqref{eq:doob-meyer-decom} for $V_{k+1}$ and $V_k$ implies
\begin{equation} \label{eq:subtraction-doob-meyer}
    V_{k+1} - V_k = M_{k+1} - M_k - A_{k+1} + A_k.
\end{equation}
It follows that
\begin{equation*}
    V_{k+1} = V_k + M_{k+1} - M_k - A_{k+1} + A_k.
\end{equation*}
Taking conditional expectation, using the martingale property of $(M_k)_{k=0}^N$ and the predictability of $(A_k)_{k=0}^N$, we obtain
\begin{equation*}
    \bE[V_{k+1} | X_k] = V_k - A_{k+1} + A_k.
\end{equation*}
Tower property implies
\begin{equation*}
    \bE[V_{k+1} | X_k] = \bE[g(\tau_{k+1}, X_{\tau_{k+1}}) | X_k] = c(t_k, X_k).
\end{equation*}
Combining the two estimates above, we derive
\begin{align*}
 V_k - A_{k+1} + A_k=c(t_k, X_k).
\end{align*}
This, together with \eqref{eq:subtraction-doob-meyer}, leads to
\begin{equation} \label{eq:dm-expression1}
    M_{k+1} - M_k = V_{k+1} - (V_k - A_{k+1} + A_k) = V_{k+1} - c(t_k, X_k).
\end{equation}
Applying a similar approach for deriving \eqref{eq:cv-linear-bsde}, we obtain
\begin{equation} \label{eq:dm-expression2}
    V_{k+1} - c(t_k, X_k) = \int_{t_k}^{t_{k+1}} \nabla_x c(s, X_s)^\top \sigma(X_s)\,\dd W_s.
\end{equation}
Finally, combining \eqref{eq:dm-expression1} and \eqref{eq:dm-expression2} yields
\begin{equation*}
    M_k = M_0 + \sum_{j=0}^{k-1} (M_{j+1} - M_j) = \sum_{j=0}^{k-1} \int_{t_j}^{t_{j+1}} \nabla_x c(s, X_s)^\top \sigma(X_s)\,\dd W_s = \int_0^{t_k} \nabla_x c(s, X_s)^\top \sigma(X_s)\,\dd W_s.
\end{equation*}
\end{proof}

\subsection{Proof of Proposition \ref{prop:value-high-bias}}
\begin{proof}
First, the unbiasedness of the schemes $\mathcal{E}_{k+1}$ and $\mathcal{E}_{k}$ implies
\begin{equation} \label{eq:unbiased-approx}
    \bE[\hat{C}_{k+1}| X_{k+1}] = \bE[V_{k+2}|X_{k+1}] \quad \text{ and }\quad \bE[\hat{C}_k| X_k] = \bE[\hat{V}_{k+1} | X_k].
\end{equation}
Next, we consider the bias of $\hat{V}_{k+1}$. Using Jensen's inequality,
\begin{align*}
\bE[\hat{V}_{k+1}|X_{k+1}] &= \bE\left[ \max\left( g(t_{k+1}, X_{k+1}), \hat{C}_{k+1}\right) | X_{k+1} \right] \\
&> \max \left( \bE[g(t_{k+1}, X_{k+1})|X_{k+1}], \bE[\hat{C}_{k+1}|X_{k+1}] \right) \\
&= \max\left( g(t_{k+1}, X_{k+1}), \mathbb{E}[V_{k+2}|X_{k+1}] \right),
\end{align*}
where the last equality follows from \eqref{eq:unbiased-approx}. Applying tower property leads to
\begin{align*}
\bE[\hat{V}_{k+1} | X_k] &= \bE[\bE[\hat{V}_{k+1}|X_{k+1}]|X_k] \\
&> \bE[ \max\left( g(t_{k+1}, X_{k+1}), \mathbb{E}[V_{k+2}|X_{k+1}] \right) |X_k] = C_k.
\end{align*}
Using \eqref{eq:unbiased-approx} again, we have
\begin{equation*}
    \bE[\hat{C}_k|X_k] > C_k,
\end{equation*}
which implies the desired conclusion.
\end{proof}

\subsection{Proof of Proposition \ref{prop:stopping-time-low-bias}}
\begin{proof}
First, the unbiasedness of the scheme $\mathcal{E}_k$ indicates
\begin{equation} \label{eq:unbiased-scheme-stopping-time}
    \bE[\tilde{C}_k | X_k] = \bE[g(\tilde{\tau}_{k+1}, X_{\tilde{\tau}_{k+1}})|X_k].
\end{equation}
Since the exact stopping time $\tau_{k+1}$ solves the subproblem $\sup_{\tau\in \mathcal{J}, \tau \ge t_{k+1}}\bE[g(\tau, X_\tau)|X_{k+1}]$, then we obtain 
\begin{equation*}
    \bE[g(\tilde{\tau}_{k+1}, X_{\tilde{\tau}_{k+1}}) | X_{k+1}] \le \bE[g(\tau_{k+1}, X_{\tau_{k+1}}) | X_{k+1}],
\end{equation*}
where the equality holds when $\tilde{\tau}_{k+1}$ is optimal. Without loss of generality, we assume the computed $\tilde{\tau}_{k+1}$ is not exact subject to small approximation error. Then, applying tower property leads to
\begin{equation*}
     \bE[g(\tilde{\tau}_{k+1}, X_{\tilde{\tau}_{k+1}}) | X_k] < \bE[g(\tau_{k+1}, X_{\tau_{k+1}}) | X_k].
\end{equation*}
Using \eqref{eq:unbiased-scheme-stopping-time} and \eqref{eq:exact-Ck-stopping-time}, we obtain
\begin{equation*}
    \bE[\tilde{C}_k | X_k] < C_k,
\end{equation*}
which implies the desired conclusion.
\end{proof}

\section{Hyperparameters and details for experiments} \label{append:hyperparameters}
Table \ref{tab:hyperparameters_ex1} reports the values of hyperparameters used in Section \ref{sec:num}. We adopt the technique of batch normalization in all subnetworks, right after the affine transformation and before activation. We double the number of epochs in the time step $t_{N-1}$ as explained in Section \ref{subsec:nn-architecture-training}. We apply a decaying learning rate for the PyTorch Adam optimizer to improve efficiency. Let $n\in \mathbb{N}$ be the training step number. In the time step $t_{N-1}$, we use the learning rate $0.01(0.0001)^{\max((n-50)/1000, 0)}$ for experiments in Section \ref{subsec:ex1-geobaskcall}, \ref{subsec:ex2-strangle}, \ref{subsec:ex3-put-heston}, and experiments of $d=2,5$ in Section \ref{subsec:ex4-max-call}, while the experiments of $d=10$ in Section \ref{subsec:ex4-max-call}, we use $0.1(0.0001)^{\max((n-50)/1000, 0)}$. In time steps $t_k$ with $k = 1,\dots, N-2$, we use the learning rate $0.01(0.0001)^{\max((n-50)/500, 0)}$ for all experiments.

\begin{table}[htbp!]
    \centering
    \begin{tabular}{ccccc c cc}
        \hline
         & $d$ & batch size & width of hidden & $N_{\rm steps}$ & & \multicolumn{2}{c}{$N_{\rm epochs}$}  \\
        \cline{7-8}
         & &  & layers $d_{\ell}$ &  && $t_{N-1}$ & $(t_k)_{k=1}^{N-2}$   \\
        \hline
        \multirow{4}{*}{Section \ref{subsec:ex1-geobaskcall}} & 3 & 8192 & 32 & 300 && 2 & 1 \\
        & 20 & 8192 & 64 & 300 && 2 & 1 \\
        & 100 & 8192 & 128 & 100 && 6 & 3 \\
        & 200 & 4096 & 128 & 100 && 8 & 4 \\
        Section \ref{subsec:ex2-strangle} & 5 & 8192 & 64 & 400 && 2 & 1 \\
        Section \ref{subsec:ex3-put-heston} & 2 & 8192 & 64 & 200 && 2 & 1 \\
        Section \ref{subsec:ex4-max-call} & 2, 5, 10 & 8192 & 64 & 400 && 2 & 1 \\
        \hline
    \end{tabular}
    \caption{Hyperparameters for experiments.}
    \label{tab:hyperparameters_ex1}
\end{table}

The calculation of projected delta values in Figure \ref{fig:delta_geobaskcall} is mainly obtained by the chain rule. First, there holds $\frac{\partial g(t_k, X_k)}{\partial \overline{X}_k} = 1$ in this example. Then combining the identity $\ln(\overline{X}_k) = \frac{1}{d}\sum_{j=1}^d \ln(X_{k,j})$ with the definition of directional derivative, we further deduce $\frac{\partial c(t_k, X_k)}{\partial \ln(\overline{X}_k)} = \sum_{j=1}^d \frac{\partial c(t_k, X_k)}{\partial \ln(X_{k,j})}$. Together with the identities $\frac{\partial c(t_k, X_k)}{\partial \ln(X_{k,j})} = \frac{\partial c(t_k, X_k)}{\partial X_{k,j}}X_{k,j}$ and $\frac{\partial c(t_k, X_k)}{\partial \ln(\overline{X}_k)} = \frac{\partial c(t_k, X_k)}{\partial\overline{X}_k}\overline{X}_k$, we finally derive
\begin{equation*}
    \frac{\partial c(t_k, X_k)}{\partial\overline{X}_k} = \frac{1}{\overline{X}_k}\sum_{j=1}^d \frac{\partial c(t_k, X_k)}{\partial X_{k,j}}X_{k,j}.
\end{equation*}

\end{document}